\documentclass[usenatbib,letterpaper,hyperref]{mn2e}

\DeclareRobustCommand{\VAN}[3]{#2}
\let\VANthebibliography\thebibliography
\def\thebibliography{\DeclareRobustCommand{\VAN}[3]{##3}\VANthebibliography}

\usepackage[labelfont=bf,labelsep=period]{caption}
\usepackage[totalwidth=480pt, totalheight=680pt]{geometry}
\usepackage{hyperref}
\usepackage{times}
\usepackage{epsfig}
\usepackage{amsmath}
\usepackage{amssymb}
\usepackage{url}
\usepackage{aas_macros}
\def\eqsim{\mathrel{\raise0.35ex\hbox{$\scriptstyle =$}\kern-0.6em
    \lower0.40ex\hbox{{$\scriptstyle \sim$}}}}
\def\gtrsim{\mathrel{\raise0.35ex\hbox{$\scriptstyle >$}\kern-0.6em
    \lower0.40ex\hbox{{$\scriptstyle \sim$}}}}
\def\lesssim{\mathrel{\raise0.35ex\hbox{$\scriptstyle <$}\kern-0.6em
    \lower0.40ex\hbox{{$\scriptstyle \sim$}}}}

\title[Uncertainty in the mass of the Milky~Way]{An overlooked source of uncertainty in the mass of the Milky~Way}
\author[K. A. Oman \& A. H. Riley]{
  Kyle A. Oman$^{1,2,3}$\thanks{kyle.a.oman@durham.ac.uk} \& Alexander H. Riley$^{1,3}$\\
  $^{1}$ Institute for Computational Cosmology, Durham University, South Road, Durham DH1 3LE, United Kingdom\\
  $^{2}$ Centre for Extragalactic Astronomy, Durham University, South Road, Durham DH1 3LE, United Kingdom\\
  $^{3}$ Department of Physics, Durham University, South Road, Durham DH1 3LE, United Kingdom\\
}
\date{\today}

\begin{document}
\label{firstpage}
\maketitle

\begin{abstract}
In the conventional approach to decomposing a rotation curve into a set of contributions from mass model components, the measurements of the rotation curve at different radii are taken to be independent. It is clear, however, that radial correlations are present in such data, for instance (but not only) because the orbital speed depends on the mass distribution at all (or, minimally, inner) radii. We adopt a very simple parametric form for a covariance matrix and constrain its parameters using Gaussian process regression. Applied to the rotation curve of the Milky~Way, this suggests the presence of correlations between neighbouring rotation curve points with amplitudes of $<10\,\mathrm{km}\,\mathrm{s}^{-1}$ over length scales of $1.5$--$2.5\,\mathrm{kpc}$ regardless of the assumed dark halo component. We show that accounting for such covariance can result in a $\sim 50$~per~cent lower total mass estimate for the Milky~Way than when it is neglected, and that the uncertainty in model parameters increases such that it seems more representative of the uncertainty in the rotation curve measurement. The statistical uncertainty associated with the covariance is comparable to or exceeds the total systematic uncertainty budget. Our findings motivate including more detailed treatment of rotation curve covariance in future analyses.
\end{abstract}

\section{Correlations across radii in rotation curve measurements}
\label{sec:intro}

Rotation curves are a widely-used dynamical tracer of the mass content of and distribution within galaxies including the Milky~Way \citep[e.g.][amongst many others]{2006ApJ...641L.109C,2008AJ....136.2648D,2008ApJ...676..920K,2009A&A...493..871S,2011MNRAS.414.2446M,2012ApJ...759..131B,2012PASJ...64...75S,2014ApJ...789...63A,2015JCAP...12..001P,2016AJ....152..157L,2019ApJ...871..120E,2020MNRAS.494.4291C,2024MNRAS.528..693O}. The atomic measurements comprising a rotation curve -- the orbital speed at fixed radius -- are usually taken to be independent from and uncorrelated with their counterparts at other radii \citep[including in all the references above; see also][for some discussion of concerns around such correlations]{2023A&A...676A.134P}. However, it is easy to see that correlations across radii must exist. One unambiguous source is the connection between gravitationally-driven kinematics and integrals of the mass distribution -- for example, the mass within some central aperture contributes to the kinematics at all larger radii, introducing a correlation. Correlations may also arise due to instrumental effects such as beam smearing \citep[][and references therein]{2009A&A...493..871S}, modelling effects such as a geometrically thick disc being imperfectly separated into rings, or physical effects such as a spiral arm coherently perturbing the kinematics over a range in radius. The presence of correlations can further be inferred by noticing that the scatter implied by the statistical uncertainties on rotation curve measurements often exceeds the point-to-point scatter measured (but misestimates of the uncertainties could also contribute, including incorrectly assuming that uncertainties are Gaussian distributed).

The heterogenous origins of radial correlations in rotation curve measurements makes them challenging to model explicitly. \citet{2022RNAAS...6..233P} proposed the pragmatic approach of assuming a parametric form for the covariance matrix:
\begin{equation}
\mathsf{K}_{ij} = a_k^2\exp\left[-\frac{1}{2}\left(\frac{\left|R_i - R_j\right|}{s_k}\right)^2\right] + \sigma_i^2\delta_{ij}.\label{eq:cov}
\end{equation}
This describes a correlation with amplitude\footnote{Our $a_k$ is equal to $\sqrt{A_k}$ in the notation of \citet{2022RNAAS...6..233P}.} $a_k$ between points of a given radial separation $R_i-R_j$ that decays exponentially with a length scale $s_k$; the uncertainties on the individual measurements are $\sigma_i$. There is freedom in the choice of `kernel function' (the first term in Eq.~(\ref{eq:cov})) but \citet{2022RNAAS...6..233P} reports that reasonable variations in the choice do not lead to large differences in results, which is enough for our illustrative purposes in this work. \citet{2022RNAAS...6..233P} argued that marginalizing over the possibility of such correlations across radii, even in such a simplistic manner, leads to more realistic and less biased confidence intervals on model parameters of interest, such as those describing the dark halo component of a galaxy.

Analysis of the rotation curve of the Milky~Way is distinct from other galaxies in two important ways. First, the systematic uncertainties affecting the measurement are distinct from those relevant in other galaxies due to our unique vantage point, especially for recent measurements incorporating high-precision proper motion measurements of stars. Second, we have more constraints on the visible matter content and distribution of our Galaxy than for external galaxies \citep[e.g. \citealp{2008gady.book.....B}, sec.~2.7,][and references therein]{2017MNRAS.465...76M,2020Galax...8...37S}, making mass models of the Milky~Way more tightly constrained. These considerations motivate us to apply the approach of \citet{2022RNAAS...6..233P}, who illustrated it using measurements of NGC~2403, to the Milky~Way in order to assess whether accounting for correlations in the measurements make up a significant portion of the uncertainty budget in mass models of the Milky Way. We also explore whether failing to account for such correlations is likely to lead to biases in rotation curve-based measurements of the mass of the Milky~Way.

\section{Mass model components and fitting methodology}
\label{sec:methods}

We use the rotation curve reported by \citet[][see their table~1]{2024MNRAS.528..693O} as a representative example of recent measurements \citep[see also][]{2023ApJ...942...12W, 2023ApJ...946...73Z, 2023A&A...678A.208J} incorporating data from the \emph{Gaia} mission \citep{2023A&A...674A...1G}. We also adopt the structural parameters for the baryonic components of the mass model of \citet[][see their table~2]{2024MNRAS.528..693O}. The circular velocity curve specified by our model given its parameters is:
\begin{equation}
  v^2_\mathrm{model}(R) = v^2_\mathrm{stars}(R) + v^2_\mathrm{gas}(R) + v^2_\mathrm{dark\,matter}(R).
\end{equation}
We have grouped together the contribution of the stellar bulge and disc (both held fixed in model optimisation) in $v_\mathrm{stars}$, and likewise the contributions of H\,\textsc{i} gas, H$_2$ gas, warm dust and cold dust in $v_{gas}$ (also all held fixed). Whereas \citet{2024MNRAS.528..693O} derived the circular velocity of each component from the enclosed mass at each radius, for disc-like components we instead use the expression in terms of the gradient of the potential $\Phi$ in the disc midplane:
\begin{equation}
  v^2=R\frac{\mathrm{d}\Phi}{\mathrm{d}R}.
\end{equation}
A derivation of the potential of a thick exponential disc can be found in \citet[][sec.~2.6.1c]{2008gady.book.....B}; we evaluate the relevant integrals numerically and use a cubic spline approximation to measure its radial derivative. For spherically-symmetric components we use the usual expression $v_\mathrm{circ}=\sqrt{GM_\mathrm{enclosed}/r}$.

We use two models for $v_\mathrm{dark\,matter}$: the \citet[][NFW]{1996ApJ...462..563N} profile and pseudo-isothermal \citep{1972ApJ...176....1G} profile. Both can be expressed as:
\begin{equation}
v^2_\mathrm{dark\,matter} = v^2_{200\mathrm{c}}\frac{R_{200\mathrm{c}}}{R}\frac{f_c\left(\frac{cR}{R_{200\mathrm{c}}}\right)}{f_c\left(c\right)},
\end{equation}
where $v^2_{200\mathrm{c}}=\frac{GM_{200\mathrm{c}}}{R_{200\mathrm{c}}}$, $R_{200\mathrm{c}}=(2GM_{200\mathrm{c}}/\Delta_\mathrm{crit}H_0^2)^\frac{1}{3}$ and $\Delta_\mathrm{crit}=200$. The mass enclosed within a sphere within which the average density is $\Delta_\mathrm{crit}$ times the critical density for closure, $M_{200\mathrm{c}}$, is a free parameter. The models differ in the definition of the second free parameter -- the `concentration paramter' $c$. For the NFW profile we adopt the usual definition $c_\mathrm{NFW}=R_{200\mathrm{c}}/R_{s}$, where $R_{s}$ is the `scale radius' in the density profile $\rho(R)\propto (R(1+R/R_s))^{-2}$. For the pseudo-isothermal profile we define $c_\mathrm{PI}=R_{200\mathrm{c}}/R_{c}$ where $R_{c}$ is the `core radius' in the density profile $\rho(R)\propto(1+(R/R_c)^2)^{-1}$. Finally, the function $f_c$ is defined for the two models respectively as\footnote{We use $\log$ and $\log_{10}$ to denote the natural and base-$10$ logarithms, respectively.}:
\begin{equation}
  f_{c,\mathrm{NFW}}(x) = \frac{\log(1+x)-x}{1+x}
\end{equation}
and
\begin{equation}
  f_{c,\mathrm{PI}}(x) = 1-\frac{\arctan(x)}{x}.
\end{equation}

We stress that the choice of these two models for $v_\mathrm{dark\,matter}$ is not motivated by their suitability to describe the Milky~Way rotation curve data. Instead, they are chosen for their similarity in terms of simplicity of mathematical form (e.g. two free parameters each, analytic expressions for all needed quantities) but stark dissimilarity in structure (central $\rho\propto r^{-1}$ and outer $\rho\propto r^{-3}$ density profile for the NFW model vs. central $\rho\propto r^{0}$ and outer $\rho\propto r^{-2}$ density profile for the pseudo-isothermal model). We will show below that some outcomes of allowing for correlations in the rotation curve data are common to both models, suggesting that the lessons learned are quite general.

We optimize the two free parameters of the model following exactly the same methodology as \citet{2022RNAAS...6..233P} -- in fact we use their software implementation, with some modification to include the stellar bulge component and the pseudo-isothermal model in addition to the NFW model, in both the case assuming independent measurements of the rotation curve at each radius and that where correlations across radii are modelled with Gaussian process (GP) regression.

\section{Impact of correlation modeling on inferred Milky Way mass profiles}
\label{sec:results}

\begin{figure*}
  \includegraphics[width=\textwidth]{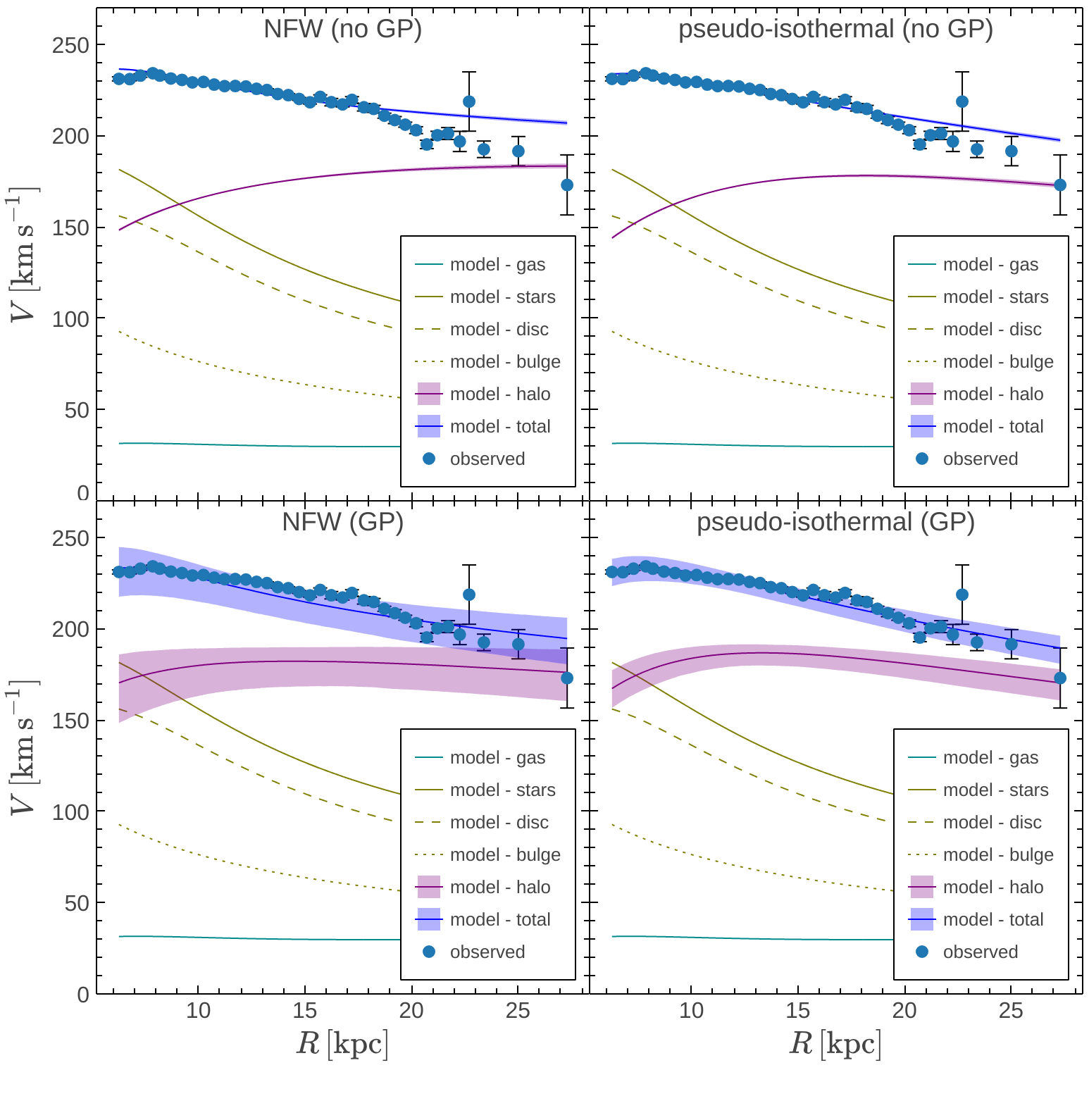}
  \caption{Illustrative mass models for the Milky Way, with and without modeling radial correlations. The measurements (points with $1\sigma$ error bars) are as presented in \citet{2024MNRAS.528..693O}. In all cases the stellar disc (dashed yellow) and bulge (dotted yellow; combined stellar components shown with solid yellow), and `gas' (green; including H\,\textsc{i} gas, H$_{2}$ gas, warm dust and cold dust) are kept fixed to the model proposed by \citet{2024MNRAS.528..693O}; see Sec.~\ref{sec:methods} for details. In the left panels the halo component (solid purple) is an NFW model, while in the right panels it is a pseudo-isothermal sphere. In the upper panels no correlation (`no GP') between the observed rotation speeds is assumed, while in the lower panels the covariance matrix for the observations is estimated using Gaussian process (`GP') regression as described by \citet{2022RNAAS...6..233P}. Shaded bands mark the regions enclosing 95~per~cent of model curves at each radius for the halo component and model total (in the upper panels these bands are very thin).}
  \label{fig:rcs}
\end{figure*}

We show the mass models resulting from our modelling for the two halo models in the cases without (upper panels) and with (lower panels) the GP regression model for radial correlations in the Milky~Way rotation curve in Fig.~\ref{fig:rcs}. Corresponding best-fitting parameter values and uncertainties are reported in Table~\ref{table:params}, and one- and two-dimensional marginalised posterior probability distributions are shown in the Appendix. In the cases without GP regression the confidence intervals on the models' total circular velocity curves are unrealistically small, and the models' inability to adequately capture the data is reflected by a reduced chi-squared ($\chi^2_\nu$) of about $10$ (NFW halo) or $5$ (pseudo-isothermal halo). Qualitatively, these fits are being driven by the measurements near $10<R/\mathrm{kpc}<15$ that have very small uncertainties and therefore miss the measurements near $20<R/\mathrm{kpc}<25$ where the uncertainties are larger.

\begin{table*}
  \caption{The first four columns show the best-fitting and marginalized $16^\mathrm{th}$-$84^\mathrm{th}$ percentile confidence intervals for the free parameters of our mass models for the two halo models -- NFW and pseudo-isothermal (P-I) -- and the case where correlations between rotation curve points are ignored (no GP) or accounted for through Gaussian process regression (GP). The last column shows the inference on the mass of the dark matter halo component of the model within a $20\,\mathrm{kpc}$ spherical aperture. Values on a linear scale are given here for ease of reference, but we note that the probability distributions are either close to log-normal ($M_{200\mathrm{c}}$, $c$, $M_\mathrm{DM}(r<20\,\mathrm{kpc})$) or asymmetric ($a_k$, $s_k$) -- see figures in the Appendix.}
  \label{table:params}
  \begin{tabular}{lccccc}
    & $M_{200\mathrm{c}}$ & & $a_k$ & $s_k$ & $M_\mathrm{DM}(r<20\,\mathrm{kpc})$\\
    model & $[10^{11}\,\mathrm{M}_\odot]$ & $c$ & $[\mathrm{km}\,\mathrm{s}^{-1}]$ & $[\mathrm{kpc}]$ & $[10^{10}\,\mathrm{M}_\odot]$ \\
    \hline
    NFW (no GP) & $8.7^{+0.2}_{-0.2}$ & $15.5^{+0.2}_{-0.2}$ & -- & -- & $15.3^{+0.1}_{-0.1}$\\
    P-I (no GP) & $3.4^{+0.1}_{-0.1}$ & $12.2^{+0.1}_{-0.1}$ & -- & -- & $14.7^{+0.1}_{-0.1}$\\
    NFW (GP) & $5.6^{+1.2}_{-1.0}$ & $25^{+6}_{-5}$ & $6^{+3}_{-2}$ & $2.3^{+0.5}_{-0.4}$ & $15.2^{+0.7}_{-0.8}$\\
    P-I (GP) & $2.8^{+0.2}_{-0.2}$& $16^{+1}_{-1}$ & $3^{+2}_{-1}$ & $1.9^{+0.6}_{-0.4}$ & $15.2^{+0.4}_{-0.5}$\\
    \hline
  \end{tabular}
\end{table*}

In the cases with GP regression the models achieve a more balanced `compromise' fit because a moderate deviation from the points with small uncertainties ($10<R/\mathrm{kpc}<15$) is mitigated by the assumption that these measurements are correlated. The correlation amplitudes and scales preferred by the models seem intuitively plausible for a galaxy like the Milky~Way: $(a_k,s_k)\sim (5\,\mathrm{km}\,\mathrm{s}^{-1}, 2.4\,\mathrm{kpc})$ for an NFW halo, or $(3\,\mathrm{km}\,\mathrm{s}^{-1}, 2.0\,\mathrm{kpc})$ for a pseudo-isothermal halo. The confidence intervals on the total circular velocity curves and dark halo components are larger and, we feel, more representative of the statistical uncertainty in the measurements. The fits are formally much better, with $\chi^2_\nu\sim 1$ for both halo models, justifying the addition of the two additional free parameters.

\begin{figure}
  \includegraphics[width=\columnwidth]{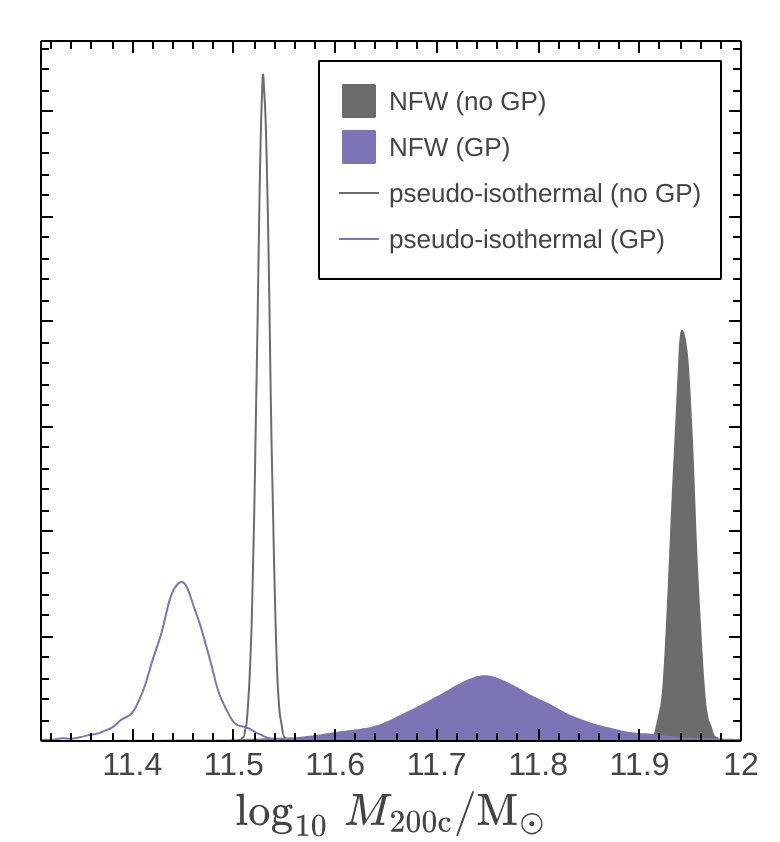}
  \caption{Marginalized posterior probability distributions for the model halo mass. The four cases shown are for an NFW halo (filled histograms) and a pseudo-isothermal halo (open histograms) for the no-correlations (`no GP'; grey) and covariance estimated with Gaussian process regression (`GP', purple) models. For both halo models, including an estimate of the coviariance results in a significant bias to lower halo mass; the posterior probability distribution also becomes wider. One- and two-dimensional marginalised posterior probability distributions for all model parameters (`corner plots') can be found in Figs.~\ref{fig:corner_nfw} \& \ref{fig:corner_iso}.}
  \label{fig:mass}
\end{figure}

In addition to wider confidence intervals, the models with GP regression are also biased towards having somewhat more centrally-concentrated dark halo components, visible in Fig.~\ref{fig:rcs} as an elevated circular velocity for the halo component near the centre ($R\lesssim 10\,\mathrm{kpc}$). This is compensated by lower halo masses such that the total dark matter mass within $16$ and $25\,\mathrm{kpc}$ is the same as in the models without GP regression for the NFW and pseudo-isothermal halo cases, respectively (the dark matter mass within $20\,\mathrm{kpc}$ for each model is tabulated in Table~\ref{table:params}). The marginalised posterior probability distributions for the halo mass for the $4$ models plotted in Fig.~\ref{fig:rcs} are shown in Fig.~\ref{fig:mass} (see also the Appendix). The variants with GP regression have systematically lower $M_{200\mathrm{c}}$, by $\sim 0.2\,\mathrm{dex}$ ($58$~per~cent) for the NFW halo or $\sim 0.1\,\mathrm{dex}$ ($26$~per~cent) for the pseudo-isothermal halo.

\section{Discussion}
\label{sec:discussion}

\subsection{Appropriateness of the halo models}

We briefly discuss whether the models shown in Fig.~\ref{fig:rcs} are realistic. The NFW halo model, in particular, comes with a strong implied prior on the concentration $c_\mathrm{NFW}$ given the mass $M_{200\mathrm{c}}$ \citep[e.g.][]{2014MNRAS.441..378L}. We have chosen not to impose this prior in our modelling, primarily to enable a fair comparison with the pseudo-isothermal halo model where no similarly strong prior exists \citep[but see][]{2004IAUS..220..377K, 2008AJ....136.2648D}. Given the existence of a mass-concentration relation for the NFW model, it makes sense to ask whether our models are consistent with it. We have deliberately not shown the relation in the $c$ versus $M_{200\mathrm{c}}$ panel of Fig.~\ref{fig:corner_nfw} -- in fact it lies largely outside of the axes. Both of our models with an NFW halo (with and without GP regression) prefer a region of the parameter space many standard deviations above the locus of the mass-concentration relation. This can reasonably be attributed to neglecting any response of the halo to the assembly of the Galaxy \citep[see e.g.][and references therein]{2020MNRAS.494.4291C}, possible mismodelling of its baryonic components, or, likely, a combination of these.

We stress, however, that our objective in this work is not to create a realistic mass model for the Milky~Way, but to highlight the kinds of systematic biases that can arise when correlations between points in the rotation curve measured at different radii are ignored. With this end in mind, the choice of halo models and how realistic they are (within reason) is irrelevant: our modelling clearly shows that biases of the same sign and similar amplitude arise in both models that we explore, despite their dissimilarity. We therefore expect that more realistic models, which likely have a central density profile somewhere between a steep NFW cusp and the flat core of the pseudo-isothermal model, suffer from the same sorts of systematic biases (e.g. in halo mass; Fig.~\ref{fig:mass}) as our models.

\begin{figure}
  \includegraphics[width=\columnwidth]{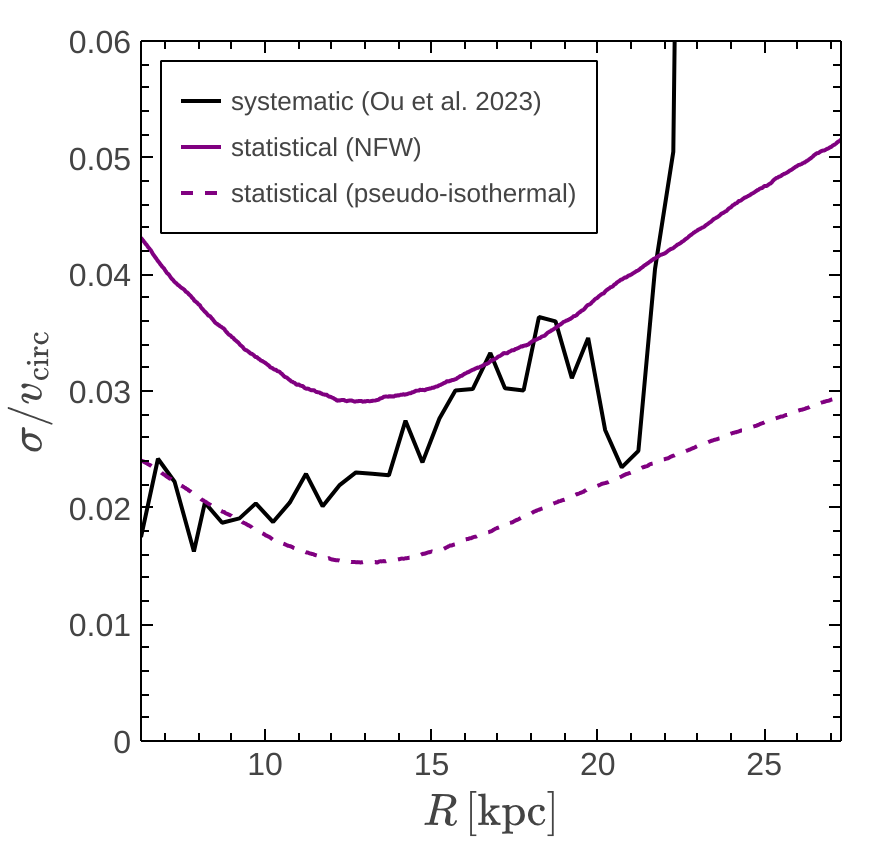}
  \caption{Comparison of the statistical uncertainty as a function of radius for our mass models including Gaussian process regression estimates of the covariance matrix. Measurements in the case with the NFW halo model (solid purple) and pseudo-isothermal halo model (dashed purple) are shown. The curves represent the width of the interval enclosing 68~per~cent of model curves at each radius, normalised by the model circular velocity at that radius. These confidence intervals are comparable in width to the estimate of the total ($1\sigma$) systematic uncertainty estimate of \citet[][black, reproduced from their fig.~6]{2024MNRAS.528..693O}.}
  \label{fig:budget}
\end{figure}

\subsection{Importance of statistical uncertainty in the uncertainty budget}

Another useful question to consider is whether the statistical uncertainty associated with correlations between circular velocity curve measurements at different radii is comparable to other leading sources of systematic uncertainty in the determination of the circular velocity curve. \citet{2024MNRAS.528..693O} provide estimates for the systematic uncertainties associated with effects such as varying the assumed density profile of the kinematic tracer population, the choice to neglect a certain higher-order term in the Jeans' equations, the uncertainty in the galactocentric Solar radius, and others in their fig.~6. We reproduce the curve showing the sum in quadrature of all of the systematic uncertainties considered in that work as a function of radius with the black line in Fig.~\ref{fig:budget}. The systematic uncertainty budget is about $2$--$3$~per~cent, with a gradual increase between about $5$ and $20\,\mathrm{kpc}$, and a sharp increase to $>10$~per~cent at larger radii.

On the same figure we show the statistical uncertainty in the total model circular velocity curve as a function of radius for our models with GP regression. Comparing to the models without GP regression (e.g. shaded bands in Fig.~\ref{fig:rcs}), it is clear that the statistical uncertainty is severely underestimated if correlations between rotation curve measurements at different radii are neglected. When correlations are accounted for, the statistical uncertainty is about $2$ to $5$~per~cent (purple solid and dashed lines for the NFW and pseudo-isothermal halo models, respectively). Whereas in the case without GP regression we would conclude that the total uncertainty budget for the model circular velocity curve is strongly systematics-dominated, in the case with GP regression it is clear that the statistical uncertainty cannot be neglected.

\subsection{Possible pitfalls in interpretation}

The approach of modelling correlations in rotation curve measurements across radii by adding degrees of freedom ($a_k$ and $s_k$) to the model and marginalizing over them requires care in interpretation. We have experimented with applying the same methodology to external galaxies and have encountered cases where the proposed mass model is unable to describe the data (an NFW dark halo being fit to a linearly rising rotation curve, for example). In such cases the parameter search responds by moving to very large values of $a_k$ (e.g. $>1000\,\mathrm{km}\,\mathrm{s}^{-1}$) and $s_k$ (e.g. $>100\,\mathrm{kpc}$), bounded above only by the boundary imposed on the (flat) prior. Such cases are clear model failures, but serve to illustrate that the parameters $a_k$ and $s_k$ will respond to essentially any discrepancy between the model and data by proposing a stronger correlation. In our exploration of the Milky~Way, it is clear that something is missing from the models because $\chi^2_\nu \gg 1$ (see Figs.~\ref{fig:corner_nfw} \& \ref{fig:corner_iso}). Allowing for correlations in the measurements provides a plausible extension to the models: the goodness of fit improves and the additional parameters converge to plausible values corresponding to a fraction of the rotation speed and linear size of the disc. The question of whether this provides a more compelling explanation than other ways of accounting for the discrepancy between the data and models when correlations are neglected remains.

It is very challenging to account for all possible sources of correlation. Attempting to write down the full covariance matrix for a rotation curve measurement is currently infeasible -- indeed this provides the motivation for the approach of \citet{2022RNAAS...6..233P}. However, taking some burden of capturing correlations off of the model by describing them in the data and its associated uncertainties would undoubtedly help to mitigate the possible pitfalls described above.

\section{Summary}
\label{sec:conclusions}

We have shown that accounting for the possibility that rotation curve measurements of the Milky~Way at different radii are statistically correlated has significant implications for inference of the Milky~Way's total mass and the structure of its dark halo. In particular:
\begin{itemize}
\item accounting for such correlations can lead to a difference in the inferred mass of the Milky~Way (lower by about 50~per~cent);
\item it also results in larger uncertainty estimates for mass model parameters that we feel are more representative of the constraining power of the data;
\item the uncertainty budget in mass models of the Milky~Way is likely not dominated by systematic uncertainties once the statistical uncertainty associated to correlations in the measurements is accounted for.
\end{itemize}

The above conclusions hold whether we assume an NFW or a pseudo-isothermal form for the dark halo component in our mass models, suggesting that they are probably generic for any broadly plausible choice of dark halo model. This strongly motivates including an allowance for correlations in the rotation curve measurements in future efforts to decompose the rotation of the Milky~Way into components, but we caution that our assumed form for the covariance matrix (Eq.~\ref{eq:cov}) is likely too simple to fully capture the correlations likely to be present in the measurements.

\section*{Acknowledgements}
\label{sec:acknowledgements}

We thank Lorenzo Posti for making a well-documented software implementation of the methods presented in \citet{2022RNAAS...6..233P} available. We thank the anonymous referee for a constructive report. KAO acknowledges support support by the Royal Society through Dorothy Hodgkin Fellowship DHF/R1/231105, by STFC through grant ST/T000244/1, and by the European Research Council (ERC) through an Advanced Investigator grant to C.S. Frenk, DMIDAS (GA 786910). AHR is supported by a Research Fellowship from the Royal Commission for the Exhibition of 1851. This research has made use of NASA's Astrophysics Data System.

\section*{Software}

This work has made use of the following software packages: \textsc{arviz} \citep{2019JOSS....4.1143K}, \textsc{astropy} \citep{2022ApJ...935..167A}, \textsc{bokeh}\footnote{\url{bokeh.pydata.org/en/latest/}}, \textsc{contourpy}\footnote{\url{pypi.org/project/contourpy/}}, \textsc{jax}\footnote{\url{github.com/google/jax}}, \textsc{numpy} \citep{2020Natur.585..357H}, \textsc{numpyro} \citep{2018arXiv181009538B,2019arXiv191211554P}, and \textsc{tinygp} \citep{2022zndo...7269074F}.

\section*{Data availability}

The rotation curve and structural parameters of the Milky~Way used as inputs in our modelling are tabulated in \citet{2024MNRAS.528..693O}, tables~1 \& 2, respectively. A software implementation of the fitting method that we have slightly adapted for this work is available from \url{https://lposti.github.io/MLPages/gaussian_processes/2022/11/02/gp_rotcurves.html}.

\bibliography{paper}

\appendix

\section{Marginalised posterior probability distributions}
\label{app:A}

\begin{figure*}
  \includegraphics[width=\textwidth]{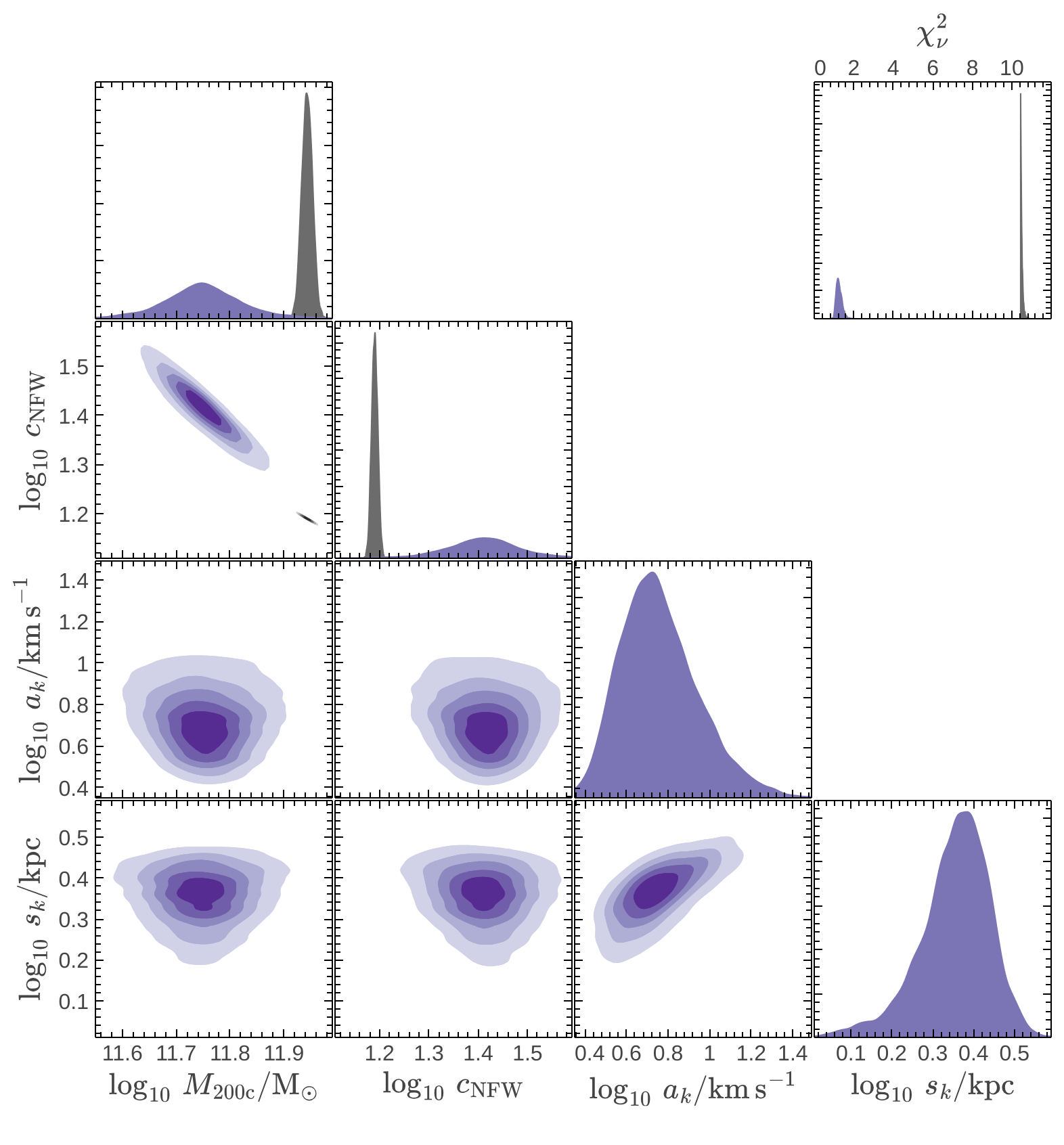}
  \caption{One- and two-dimensional marginalised posterior probability distributions for the parameters of our mass models with an NFW halo model in the `no GP' (grey) and `GP' (purple) cases. Model parameters in the `no GP' case are the halo mass $M_\mathrm{200\mathrm{c}}$ and concentration $c_\mathrm{NFW}\equiv R_{200\mathrm{c}}/R_\mathrm{s}$. The GP case supplements these with a correlation amplitude $a_k$ and length scale $s_k$. The upper right panel shows the distribution of $\chi^{2}_{\nu}$ values for samples in the Markov chains.}
  \label{fig:corner_nfw}
\end{figure*}

\begin{figure*}
  \includegraphics[width=\textwidth]{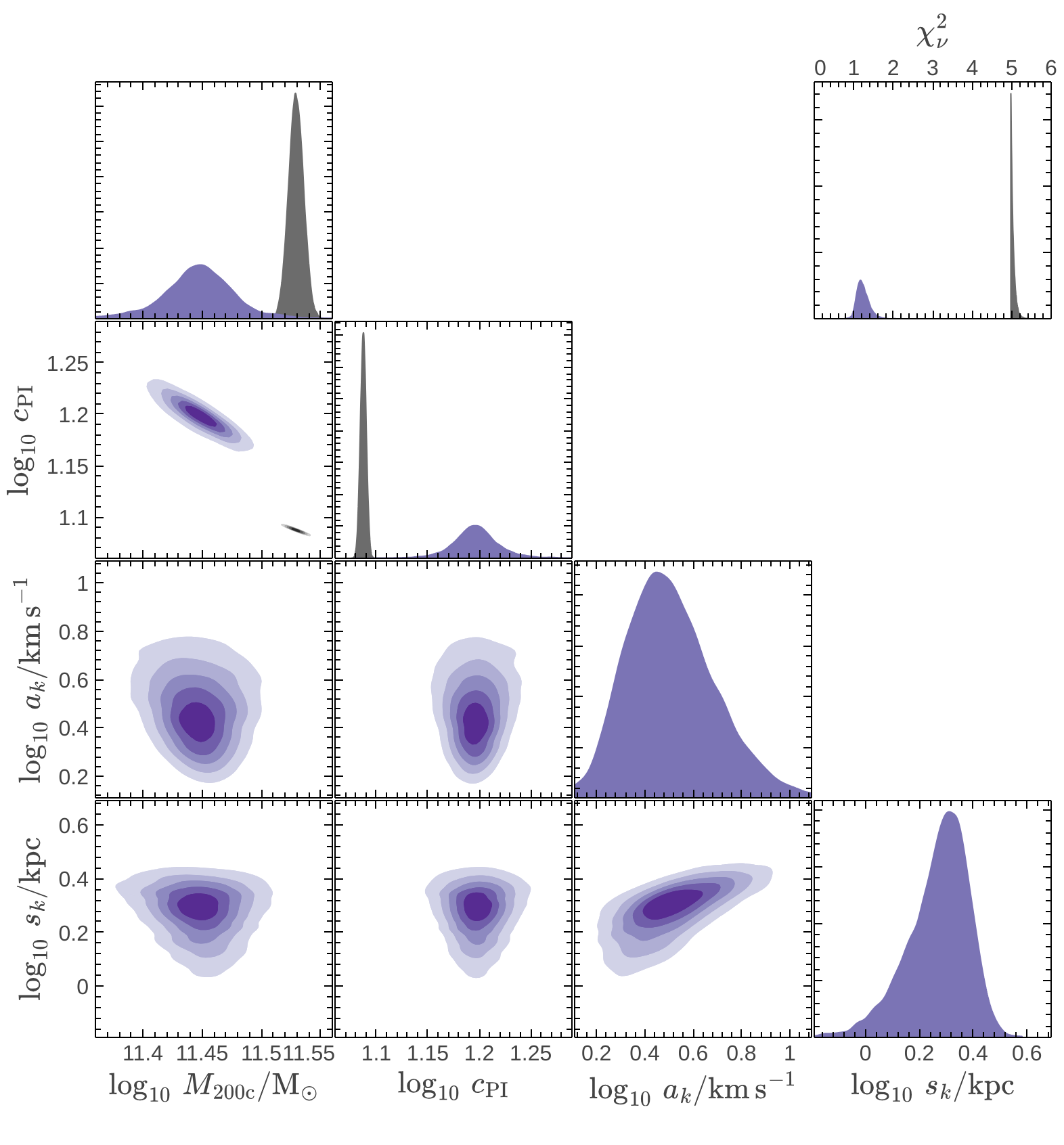}
  \caption{As Fig.~\ref{fig:corner_nfw}, but for our mass models with a pseudo-isothermal halo model. For these models the `concentration' parameter is defined $c_\mathrm{PI}\equiv R_{200\mathrm{c}}/R_\mathrm{c}$.}
  \label{fig:corner_iso}
\end{figure*}

We show the one- and two-dimensional marginalised posterior probability distributions (`corner plots') for the parameters ($M_{200\mathrm{c}}$, $c$, $a_k$ and $s_k$) of our mass models for the case including an NFW halo model in Fig.~\ref{fig:corner_nfw}, and for the case including a pseudi-isothermal halo model in Fig.~\ref{fig:corner_iso}. We also include a panel in each figure showing the distribution of $\chi^2_\nu$ values used in the likelihood function \citep[for a definition see][eq.~(2)]{2019A&A...626A..56P}.

\label{lastpage}
\end{document}